\documentclass[twocolumn,showpacs,preprintnumbers,amsmath,amssymb,nofootinbib,scrartcl]{revtex4}
\input psfig.sty
\usepackage{epsf}
\usepackage{graphicx,color}  
\usepackage{dcolumn}   
\usepackage{bm}        

\newcommand{\be}{\begin{equation}}
\newcommand{\en}{\end{equation}}
\newcommand{\bea}{\begin{eqnarray}}
\newcommand{\ena}{\end{eqnarray}}

\begin{document}

\title{Cosmological Consequences of Exponential Gravity in Palatini Formalism}

\author{M. Campista$^1${\footnote{campista@on.br}}}

\author{B. Santos${^1}${\footnote{thoven@on.br}}}

\author{J. Santos$^2${\footnote{janilo@dfte.ufrn.br}}}

\author{J. S. Alcaniz${^1}${\footnote{alcaniz@on.br}}}

\affiliation{$^1$Observat\'orio Nacional, 20921-400 Rio de Janeiro - RJ, Brasil}

\affiliation{$^2$Departamento de F\'{\i}sica, Universidade Federal do Rio G. do Norte,  59072-970 Natal - RN, Brasil}

\date{\today}


\begin{abstract}

We investigate cosmological consequences of a class of exponential $f(R)$ gravity in the Palatini formalism. By using the current largest type Ia Supernova sample along with determinations of the cosmic expansion at intermediary and high-$z$ we impose tight constraints on the model parameters. Differently from other $f(R)$ models, we find solutions of transient acceleration, in which the large-scale modification of gravity will drive the Universe to a new decelerated era in the future. We also show that a viable cosmological history with the usual matter-dominated era followed by an  accelerating phase is predicted for some intervals of model parameters.

\end{abstract}

\pacs{95.30.Sf, 04.50.Kd, 95.36.+x}

\maketitle


\section{Introduction}


Nowadays, there is great interest in modified $f(R)$ gravity (see~\cite{Francaviglia} for recent reviews). Such theories are interesting in that they generalize Einstein's general relativity and provide insights into the consequences of quantum corrections to its equations in the high energy regime ($R\rightarrow \infty$). In cosmology, the interest in these theories comes from the fact that they can  naturally drive an accelerating cosmic expansion without introducing dark energy, as happens for instance in the standard $\Lambda$CDM cosmology. However, the freedom in the choice of different functional forms of $f(R)$ gives rise to the problem of how to constrain the many possible $f(R)$ gravity theories. In this regard, much efforts have been developed so far, mainly from the theoretical viewpoint~\cite{Ferraris}. General principles such as the so-called energy conditions~\cite{energy_conditions}, nonlocal causal structure~\cite{RSantos}, have also been taken into account in
  order to clarify its subtleties. More recently, observational constraints from several cosmological data sets have been explored for testing the viability of these theories~\cite{Amarzguioui}.

An important aspect that is worth emphasizing concerns the two different variational approaches that may be followed when one works with $f(R)$ gravity theories, namely, the metric  and the Palatini formalisms (see, e.g., \cite{Francaviglia}). In the metric formalism the connections are assumed to be the Christoffel symbols and the variation of the action is taken with respect to the metric, whereas in the Palatini variational approach the metric and the affine connections are treated as independent fields and the variation is taken with respect to both. Because in the Palatini approach the connections depend on the particular $f(R)$, while in metric formalism the connections are defined {\it a priori} as the Christoffel symbols, the same $f(R)$ seems to lead to different space-time structures.

In fact, these approaches are certainly equivalents in the context of general relativity (GR), i.e., in the case of linear Hilbert action; for a general $f(R)$ term in the action, they seem to provide completely different theories, with very distinct equations of motion.   The Palatini variational approach, for instance, leads to 2nd order differential field equations, while the resulting field equations in the metric approach are 4th order coupled differential equations, which presents quite unpleasant behavior. These differences also extend to the observational aspects. For instance, we note that cosmological models based on a power-law functional form in the metric formulation fail in reproducing the standard matter-dominated era followed by an acceleration phase~\cite{Amendola} (see, however, \cite{cap}), whereas in the Palatini approach, analysis of a dynamical autonomous systems for the same Lagrangian density have shown that such theories admit the three post inflationary phases of the standard cosmolog
 ical model~\cite{Tavakol}. Although being mathematically  more simple and successful in passing cosmological tests, we do not yet have a clear comprehension of the properties of the Palatini formulation of $f(R)$ gravity in other scenarios, and issues such as solar system experiments~\cite{Faraoni}, the Newtonian limit~\cite{Meng-Wang} and the Cauchy problem~\cite{Lanahan} are still contentious (regarding this last issue, see however Refs.~\cite{Cauchy-Problem}).

In this paper, we explore cosmological consequences of a class of exponential $f(R)$-gravity models in the Palatini formalism. In order to test the observational viability of these scenarios, we use one of the latest type Ia Supernovae (SNe Ia) sample, the so-called Union2 compilation~\cite{Amanullah} along with 11 determinations of the expansion rate $H(z)$~\cite{newh,svj}. In what concerns the past evolution of the Universe, we show that for some intervals of model parameters a matter-dominated era is followed by a late time accelerating phase, differently from some results in the metric approach. Another interesting feature of this class of models is the possibility of a transient cosmic acceleration, which can lead the Universe to a new matter-dominated era in the future. This particular result seems to be in agreement with current requirements from String/M theory.


\section{Palatini $f(R)$ Cosmologies}  \label{Palatini-app}


$f(R)$-cosmologies are based on the modified Einstein equations of motion derived from the dubbed $f(R)$ gravity.
The action that defines an $f(R)$ gravity is given by
\begin{equation}
\label{actionJF}
S = \frac{1}{2\kappa^2}\int d^4x\sqrt{-g}f(R) + S_m\,,
\end{equation}
where $\kappa^2=8\pi G$, $g$ is the determinant of the metric tensor and $S_m$ is the standard action for the matter
fields. Treating the metric and the connection as completely independent fields, variation of this action with respect
to the metric provides the field equations
\begin{equation}
\label{field_eq}
f'R_{(\mu\nu)} - \frac{f}{2}g_{\mu\nu}  = \kappa^2T_{\mu\nu}\,;
\end{equation}
while variation with respect to the connection gives
\begin{equation}  \label{connections_eq}
\tilde{\nabla}_\beta\left( f'\sqrt{-g}\,g^{\mu\nu}\right)=0\,.
\end{equation}
In (\ref{field_eq}) $T_{\mu\nu}$ is the matter energy-momentum tensor which, for a perfect-fluid, is given by
$T_{\mu\nu} = (\rho_m + p_m)u_{\mu}u_{\nu} + p_m g_{\mu\nu}$,
where $\rho_m$ is the energy density, $p_m$ is the fluid pressure and $u_{\mu}$
is the fluid four-velocity. Here, we adopt the notation $f'=df/dR$ and
$f''=d^2f/dR^2$. Equation (\ref{connections_eq}) give us the connections $\Gamma_{\mu\nu}^{\rho}$,
which are related with the Christoffel symbols $\left\{^{\rho}_{\mu\nu}\right\}$ of the metric $g_{\mu\nu}$ by
\begin{equation}  \label{connections}
\Gamma_{\mu\nu}^{\rho} = \left\{^{\rho}_{\mu\nu}\right\} + \frac{1}{2f'}\left( \delta^{\rho}_{\mu}\partial_{\nu} +
\delta^{\rho}_{\nu}\partial_{\mu} - g_{\mu\nu}g^{\rho\sigma}\partial_{\sigma} \right)f'\,.
\end{equation}
Note that in (\ref{field_eq}), $R_{\mu\nu}$ must be calculated in the usual way, i.e., in terms of the independent connection
$\Gamma_{\mu\nu}^{\rho}$, given by (\ref{connections}) and its derivatives.

We assume a flat homogeneous and isotropic Friedmann-Lema\^{i}tre-Robertson-Walker universe whose metric is
$g_{\mu\nu}=diag(-1,a^2,a^2,a^2)$, where $a(t)$ is the cosmological scale factor. The generalized Friedmann equation, obtained from (\ref{field_eq}), can be written in terms of the redshift parameter $z=a_0/a -1$
and the density parameter $\Omega_{mo} \equiv \kappa^2\rho_{mo}/(3H_0^2)$ as 
\begin{equation}
\label{fe3}
{H} = H_0\left[\frac{3\Omega_{mo}(1 + z)^3 + f/H_0^2}{6f'\left(
1 + \frac{9}{2}\,\frac{f''}{f'}\,\frac{H_0^2\Omega_{mo}(1+z)^3}{Rf'' - f'}\right)^2}\right]^{1/2},
\end{equation}
where $\rho_{mo}$ is the matter density today. In terms of these quantities, the trace of Eq. (\ref{field_eq}) also provides an important relation:
\begin{equation}
\label{trace2}
Rf' - 2f = -3H_0^2\Omega_{mo}(1 + z)^3\,.
\end{equation}

\subsection{The Gravity Model}

There has been an increasing recent interest in models of exponential gravity (see, e.g., \cite{Cognola,Linder,Bamba,yang} and references therein).
Here, we explore a theory of the type:
\begin{equation}  \label{expo-gravity}
f(R) = R - \alpha nH_0^2\left( 1- e^{-R/\alpha H_0^2} \right)\,,
\end{equation}
in the Palatini formalism. In the above expression,  $n$ is a free parameter of the theory  and $\alpha$ corresponds to the strength to which the curvature $R$ scales with the Hubble parameter $H_0$. A functional form of this type was originally studied in the metric formalism by Cognola et al. \cite{Cognola} and Linder \cite{Linder}, and some viability conditions of this model  were investigated  in Refs.~\cite{Bamba,yang}.  
In the next section, we discuss some cosmological consequences of the exponential gravity theory given by Eq. (\ref{expo-gravity}). To perform our analysis, we impose the positivity of the effective gravitational coupling $\kappa^2/f'(R)>0$ (to avoid anti-gravity and to guarantee that the graviton is not a ghost in the sense of a quantum theory), which leads to the constraint $n <\exp{(R/\alpha H_0^2)}$.

\section{Observational constraints}

Figures 1 and 2 show, respectively, the evolution of the Hubble parameter (Eq. \ref{fe3}) and the predicted distance modulus $\mu{(z)} = 5\log[d_L(z)]+ 25$ as a function of  redshift for some best-fit values for $n$, $\alpha$ and $\Omega_m$ discussed in this paper. In the latter expression, $d_L = (1+z)\int_{0}^{z}\frac{dz^{\prime}}{H(z^{\prime})}$ stands for the luminosity distance (in units of megaparsecs and $c = 1$). For the sake of comparison, the standard $\Lambda$CDM prediction with $\Omega_m = 0.27$ is also shown (thick line). Note that all models seem to be able to reproduce fairly well both the $H(z)$ and  SNe Ia measurements.

\subsection{Hubble Evolution}

The data points in Fig. 1 are $H(z)$ determinations taken from Ref.~\cite{newh,svj}. These determinations are based on the differential age method that relates the Hubble parameter  directly to the measurable quantity $dt/dz$ by $H(z) = -\dot{z}/(1+z)$~\cite{Jimenez}, and can be achieved from the recently released sample of old passive galaxies from Gemini Deep Deep Survey (GDDS)~\cite{gemini} and archival data~\cite{archival}\footnote{The same data, along with other age estimates of high-$z$ objects, were recently used to reconstruct the shape and redshift evolution of the dark energy potential~\cite{svj}, to place bounds on holography-inspired dark energy scenarios~\cite{ap1}, as well as to impose constraints on some classes of $f(R)$ models~\cite{ap2}.}.

\begin{figure}[t]
	\centerline{ \psfig{figure=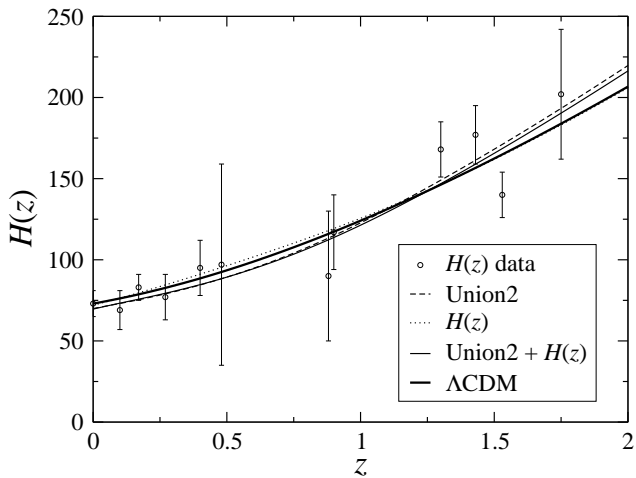,width=3.45truein,angle=-90} \hskip 0.1in }
	\caption{ The predicted Hubble evolution $H(z)$ as a function of the redshift for the exponential gravity model given by Eq. (\ref{expo-gravity}). The curves correspond to the best-fit values of $\Omega_m$ $n$ and $\alpha$ obtained from $H(z)$, Union2 and Union2 + $H(z)$ data analyses discussed in the text. For the sake of comparison, the standard $\Lambda$CDM model prediction is also shown. The data points are the measurements of the $H(z)$ given in Ref.~\cite{svj}. }
	\label{figh}
\end{figure}

In order to impose quantitative constraints from these $H(z)$ data on models of exponential gravity, as given by Eq. (\ref{expo-gravity}), we minimize the function
\begin{equation}
	\chi_H^2 = \sum_{i=1}^{N_H} \frac{ \left[ H_{th}^i (z|\mathbf{P}) - H_{obs}^i (z) \right]^2 }{ \sigma^2_i } \;.
\end{equation}
In the above expresion, $H_{th}^i(z|\mathbf{P})$ is the theoretical Hubble parameter at redshift $z_i$, which depends on the complete set of parameters $\mathbf{P} \equiv (\Omega_{m}, n, \alpha)$, $H_{obs}^i(z)$ stands for the values of the Hubble parameter given in Ref.~\cite{newh} and $\sigma_i$ is the uncertainty for each of the $N_H = 12$ determinations of $H(z)$. In our analysis, we added to this $H(z)$ sample a recent estimate of the current value of the Hubble parameter, $H_0 = 72 \pm 8$ ${\rm{km.s^{-1}.Mpc^{-1}}}$, as given by the final results of the HST key project~\cite{h0}.

Table I shows the results of our statistical analysis. From the above  $\chi^{2}_{H}$ function we construct a likelihood function ${\cal{L}} \propto \exp({-\chi_{H}^2/2})$ and derive the $1, 2$ and $3\sigma$ intervals  for the parameters $\Omega_{m}$, $n$ and $\alpha$. The best-fit parameters are the values $\mathbf{\bar{P}}$ that maximize ${\cal{L}}$ and the $1, 2$ and 3$\sigma$ confidence intervals are defined as the sets of cosmological parameters $\mathbf{P}_{\sigma}$ at which the likelihood ${\cal{L}}(\mathbf{P}_{\sigma})$ is $\exp(-1/2)$, $\exp(-4/2)$ and $\exp(-9/2)$ times smaller than the maximum likelihood ${\cal{L}(\mathbf{\bar{P}})}$.  From this $H(z)$ analysis, the best-fit values found are  $\Omega_{m} = 0.27$, $n = 1.56$ and $\alpha = 2.56$.  As expected, due to the current large uncertainties on the $H(z)$ measurements, we clearly see that these data alone do not tightly constrain the values of $n$ and $\alpha$.

\begin{figure}[t]
	\centerline{ \psfig{figure=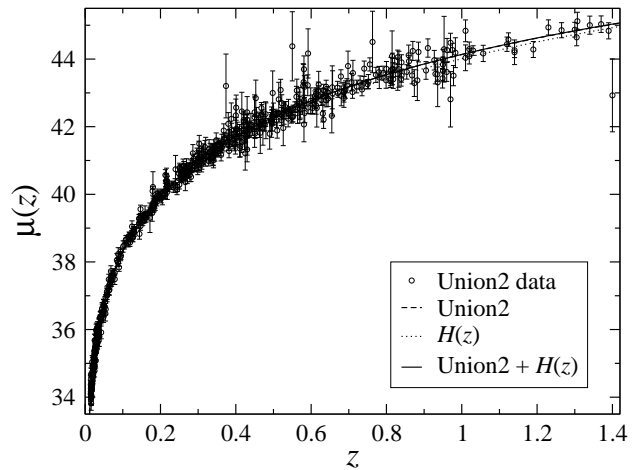,width=3.45truein,angle=-90} \hskip 0.1in }
	\caption{ Hubble diagram for 557 SNe Ia from the Union2 sample~\cite{Amanullah}. The curves correspond to the best-fit values of $\Omega_m$ $n$ and $\alpha$ arising from statistical analyses involving $H(z)$, SNe Ia and $H(z)$ +SNe Ia. }
	\label{figh}
\end{figure}

\begin{table*}[]
		\small
		\begin{center}
                        \caption{$1, 2$ and $3\sigma$ intervals for $f(R)$ parameters}
			\begin{tabular}{l|c|c|c|c}
				\toprule
				\toprule
						& $\Omega_{m0}$										& $n$											& $\alpha$							& $\chi^2_{\text{min}}/\nu$ \\
				\hline
				\hline \\
                                $H(z)$		& $0.27_{-0.03}^{+0.03}(1\sigma)_{-0.06}^{+0.07}(2\sigma)_{-0.09}^{+0.10}(3\sigma)$	& $1.56_{-0.56}^{+0.41}(1\sigma)_{-1.14}^{+0.77}(2\sigma)_{-1.21}^{+1.13}(3\sigma)$	& $2.56_{-0.49}^{+0.53}(1\sigma)_{-0.94}^{+1.09}(2\sigma)_{-1.34}^{+1.69}(3\sigma)$	& 0.84 \\
				&&&&\\
				Union2		& $0.33_{-0.003}^{+0.005}(1\sigma)_{-0.01}^{+0.01}(2\sigma)_{-0.02}^{+0.02}(3\sigma)$	& $0.60_{-0.01}^{+0.23}(1\sigma)_{-0.02}^{+0.26}(2\sigma)_{-0.04}^{+0.28}(3\sigma)$	& $1.48_{-0.01}^{+0.02}(1\sigma)_{-0.03}^{+0.03}(2\sigma)_{-0.04}^{+0.05}(3\sigma)$	& 0.97 \\
				&&&&\\
				Union2 + $H(z)$	& $0.32_{-0.003}^{+0.004}(1\sigma)_{-0.01}^{+0.01}(2\sigma)_{-0.01}^{+0.02}(3\sigma)$	& $0.56_{-0.003}^{+0.007}(1\sigma)_{-0.01}^{+0.36}(2\sigma)_{-0.02}^{+0.38}(3\sigma)$	& $1.56_{-0.01}^{+0.02}(1\sigma)_{-0.03}^{+0.04}(2\sigma)_{-0.05}^{+0.06}(3\sigma)$	& 0.97 \\
				\hline
				\hline
			\end{tabular}
		\end{center}
	\end{table*}

\subsection{SNe Ia}

The number and quality of SNe Ia data available for cosmological studies have increased considerably in the past few years. One of the most up-to-date SNe Ia data sets has been compiled by Amanullah {\it et al.} \cite{Amanullah}, the so-called Union2 sample. This sample is an update of the original Union compilation that comprises 557 data points including recent large samples from other surveys and uses SALT2 for SN Ia light-curve fitting.

Similarly to the $H(z)$ test, we estimated the best-fit to the set of parameters $\mathbf{P}$ by using a $\chi^{2}$ statistics, with
\begin{equation} \label{chi2307}
\chi^{2}_{SNe} = \sum_{i=1}^{N_{SNe}}{\frac{\left[\mu_0^{i}(z|\mathbf{P}) -
\mu_{obs}^{i}(z)\right]^{2}}{\sigma_i^{2}}},
\end{equation}
where $\mu_p^{i}(z|\mathbf{P})$ is the predicted distance modulus given above, $\mu_o^{i}(z)$ is the extinction corrected distance modulus for a given SNe Ia at $z_i$, and $\sigma_i$ is the uncertainty in the individual distance moduli. Since we use in our analyses the Union2 sample (see \cite{Amanullah} for details),  $N_{SNe} = 557$.

1, 2 and  $3\sigma$ intervals for the parameters $\Omega_{m}$, $n$ and $\alpha$ are shown in Table I. For this SNe Ia analysis the best-fit values are $\Omega_{m} = 0.33$, $n = 0.60$ and $\alpha = 1.48$ with $\chi^2_{min}/\nu \simeq 0.97$ ($\nu$ stands for the number of degree of freedom). Note that the intervals are now considerably tighter than those obtained from the $H(z)$ analysis described above, which reflects the greater constraining power of SNe Ia data when compared with the current $H(z)$ sample.

\begin{figure}[t]
\centerline{\psfig{figure=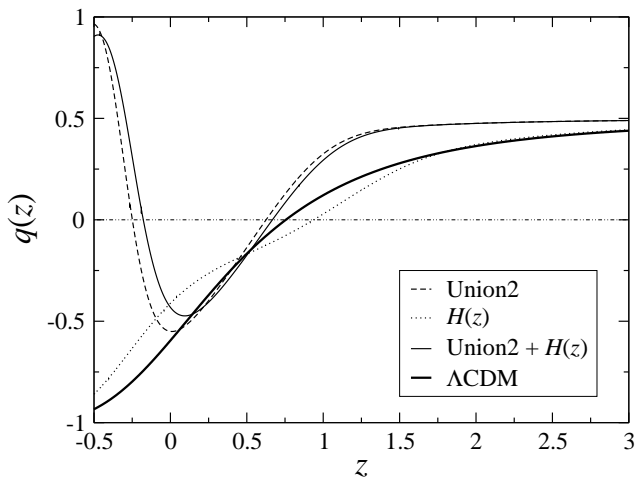,width=3.45truein,angle=-90}
\hskip 0.1in}
\caption{Deceleration parameter as a function of $z$ for the best-fit values of $\Omega_m$ $n$ and $\alpha$ presented in Table I. Note that for some combinations of parameters the cosmic acceleration is a transient phenomenon. The $\Lambda$CDM model, whose predicted cosmic acceleration is eternal, is also shown for the sake of comparison.}
\label{figh}
\end{figure}

\begin{figure}[t]
\centerline{\psfig{figure=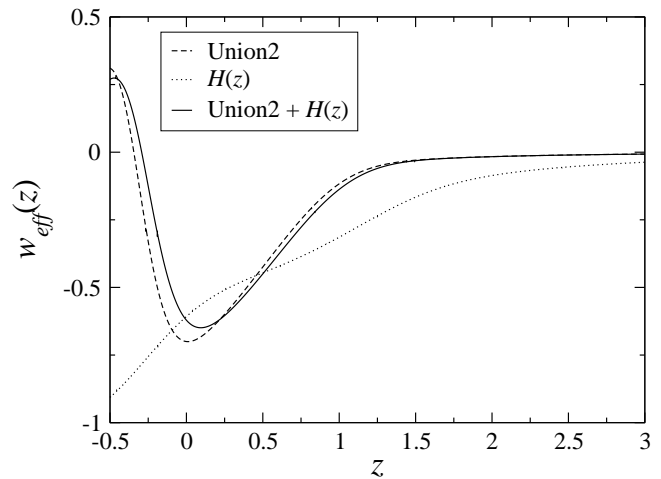,width=3.45truein,angle=-90}
\hskip 0.1in}
\caption{ Effective equation of state as a function of redshift for the exponential gravity theory of Eq. (\ref{expo-gravity}) in the Palatini formalism. As in the previous figures, the curves correspond to the best-fit values of $\Omega_m$ $n$ and $\alpha$ obtained from $H(z)$, Union2 and Union2 + $H(z)$ data analyses and shown in Table I.}
\label{figh}
\end{figure}

For completeness, we also performed a joint analysis by considering $\chi^2_T = \chi_H^2 + \chi_{SNe}^2$. The best-fit values for this analysis are $\Omega_{m} = 0.32$, $n = 0.56$ and $\alpha = 1.56$ with $\chi^2_{min}/\nu \simeq 1.0$. At $2\sigma$ level, we found $0.31 \leq \Omega_{m} \leq 0.33$, $0.55 \leq n \leq 0.92$ and $1.53 \leq \alpha \leq 1.60$ (see also Table I).

\section{Cosmological consequences}

\subsection{Acceleration history}

Refs.~\cite{fischler} have seriously pointed out a possible conflict between an eternally accelerating universe and our best candidate for a consistent quantum theory of gravity, i.e., String/M theories. The reason is that the only known formulation of String theory is in terms of S-matrices, which require infinitely separated, noninteracting  in and out states. As is well known, in the standard $\Lambda$CDM scenario the universe will asymptotically become a de-Sitter space, which has a cosmological event horizon ($\Delta_{\rm{h}} = \int{da/a^2 H(a)}$) with physics confined to a finite region and, therefore, no isolated states.

In this regard, an interesting feature of the exponential $f(R)$ gravity discussed above is the possibility of a transient cosmic acceleration with $\Delta_{\rm{h}} \rightarrow \infty$. To study this phenomenon, let us consider the deceleration parameter
\begin{equation}
q(z) = \frac{(1+ z)}{H(z)}H'(z) - 1\;,
\end{equation}
where a prime denotes differentiation with respect to $z$ and $H(z)$ is given by Eq. (\ref{fe3}).

Figure 3 shows $q(z)$ as a function of the redshift for the three sets of best-fit values obtained in the statistical analyses of Sec. III. As can be seen from this figure, for some combinations of parameters the Universe was decelerated in the past, switched to the current accelerating phase at $z_{a} \simeq 1$ and will eventually decelerate again at some $z_{d} < -1$. For these sets of parameters, it is possible to show that $\Delta_{\rm{h}} \rightarrow \infty$, thereby alleviating the potential theoretical and observational conflict discussed above. It is worth emphasizing that this kind of dynamic behavior is not found in most of the $f(R)$ cosmologies discussed in the literature~\cite{janilo}, being essentially a feature of the so-called thawing and hybrid quintessence potentials~\cite{hybrid}, some classes of coupled quintessence models~\cite{ernandes} and  brane-world scenarios~\cite{sahni}.


\subsection{Effective equation of state}

In Ref.~\cite{Amendola}, it was shown that $f(R)$ derived cosmologies in the metric formalism cannot produce a standard matter-dominated era followed by an accelerating expansion (we refer the reader to \cite{cap} for a different conclusion). To verify if this undesirable behavior happens in the Palatini $f(R)$ gravity discussed in this paper, we derive the effective equation of state (EoS)
\begin{equation}
w_{eff} = -1 + \frac{2(1+z)}{3H}H'(z)
\end{equation}
as a function of the redshift.

In Figure 4, we  show the effective EoS as a function of $z$ for the best-fit values discussed in the previous section. Note that, for some of these combinations of parameters (basically those derived from SNe Ia data), the universe went through a past matter-dominated phase ($w=0$) before switching to a late time accelerating phase ($w < 0$). In particular, we note that for the best-fit values derived from the SNe Ia plus $H(z)$ joint analysis, there seems to be evidence for a slowing down of the cosmic acceleration today, which is somewhat in agreement with the results of Ref.~\cite{sahni1} for some dark energy parameterizations. From the results shown in Fig. 5, we clearly see that the arguments of Ref.~\cite{Amendola} about the behavior of $w_{eff}$ in the metric approach seems not to apply to the Palatini formalism, at least for the exponential $f(R)$ gravity theory studied here and the interval of parameters $\Omega_{m}$, $n$ and $\alpha$ given by our statistical analyses.

\vspace{0.5cm}


\section{Concluding Remarks}  \label{Conclusion}


Cosmological models based on $f(R)$-gravity may exhibit a natural acceleration mechanism without introducing a dark energy component. In this paper, we have investigated cosmological consequences of a class of exponential $f(R)$-gravity in the Palatini formalism, as given by Eq. (\ref{expo-gravity}). We have performed consistency checks and tested the observational viability of these scenarios by using one of the latest SNe Ia data, the so-called Union2 sample with 557 data points and 11 measurements of the expansion rate $H(z)$ at intermediary and high-$z$. We have found a good agreement between these observations and the theoretical predictions of the model, with the reduced $\chi^2_{min}/\nu \simeq 1$ for the three tests performed.

Differently from the dynamical behavior of other $f(R)$ scenarios discussed in the literature (either in metric or Palatini formalisms), we have found solutions of transient cosmic acceleration in which the large-scale modification of gravity will drive the Universe to a new matter-dominated era in the future. As mentioned earlier, this kind of solution is in full agreement with theoretical requeriments from String/M theories, as first pointed out in Ref.~\cite{fischler}.

Finally, we have also shown that, differently from the results of Ref.~\cite{Amendola} for power-law $f(R)$ gravity in the  metric formalism, exponential $f(R)$ models corresponding to the best-fit solutions from SNe Ia and SNe Ia + $H(z)$ $\chi^2$ minimization have the usual matter-dominated phase followed by a late time cosmic acceleration (see also \cite{cap} for a discussion).

\begin{acknowledgments}

The authors acknowledge financial support from CNPq and CAPES.

\end{acknowledgments}

\end{document}